# LUMPY: A probabilistic framework for structural variant discovery

Ryan M Layer[1], Aaron R Quinlan*[1,2,3,4] and Ira M Hall*[2,4]

[1]Department of Computer Science

[2]Department of Biochemistry and Molecular Genetics

[3]Department of Public Health Sciences

[4]Center for Public Health Genomics, University of Virginia, Charlottesville, VA

*Corresponding authors

Please address correspondence to:

Ira M. Hall

Department of Biochemistry and Molecular Genetics

Center for Public Health Genomics

University of Virginia

Charlottesville, VA 22908

USA

irahall@virginia.edu

Aaron R. Quinlan

Center for Public Health Genomics

Department of Public Health Science

Department of Biochemistry and Molecular Genetics

Department of Computer Science

University of Virginia

Charlottesville, VA 22908

USA

arq5x@virginia.edu

# Abstract

**Background:** Comprehensive discovery of human genome structural variation (SV) from DNA sequencing data requires the integration of multiple read alignment signals including read-pair, split-read and read-depth. However, owing to inherent technical challenges, existing SV discovery algorithms either use a single signal in isolation, or at best use two signals sequentially. Such approaches suffer from limited sensitivity when a variant is captured by few reads, as often occurs in low coverage datasets or at somatic mutations with low intra-sample allele frequencies. In particular, sensitive detection of low frequency somatic SVs in heterogeneous tumor samples is an unsolved problem with clinical implications.

**Results:** We present LUMPY, a novel and extremely flexible probabilistic SV discovery framework capable of integrating any number of SV detection signals – including those generated from read alignments or prior evidence – jointly across multiple samples. LUMPY enables straightforward signal integration by mapping each SV signal to a common abstract representation in the form of breakpoint probability distributions, and then performs SV prediction operations at this higher level. This approach allows for simple and natural signal integration, inherently produces a probabilistic measure of breakpoint position, and can be easily extended to new signals as sequencing technologies and SV detection strategies evolve. We demonstrate that integration of read-pair and split-read signals yields improved sensitivity over extant methods, especially when SV signal is reduced owing to either low coverage or low variant frequency in heterogeneous samples such as cancer.

**Conclusions:** SV detection sensitivity is greatly improved when multiple alignment signals are examined simultaneously in a single probabilistic model. This advance will be especially useful for applications that require detection of variants from low coverage data or heterogeneous tumors.

# Background

Differences in chromosome structure are a prominent source of human genetic variation. These differences are collectively known as structural variation (SV), a term that encompasses diverse genomic alterations including deletion, duplication, insertion, inversion, translocation or complex rearrangement of relatively large (e.g., >100 bp) segments. While SVs are considerably less common than smaller-scale forms of genetic variation such as single nucleotide polymorphisms (SNPs), they have greater functional potential due to their larger size, and they are more likely to alter gene structure or dosage.

Our current understanding of the prevalence and impact of SV has been driven by recent advances in genome sequencing. However, the discovery and genotyping of SV from DNA sequence data has lagged far behind SNPs because it is fundamentally more difficult. SVs vary considerably in size, architecture and genomic context, and read alignment accuracy is compromised near SVs by the presence of novel junctions (i.e., breakpoints) between the "sample" and reference genomes. Moreover, SVs generate multiple alignment signals including altered sequence coverage within duplications or deletions (read-depth), breakpoint-spanning paired-end reads that align discordantly relative to each other (read-pair), and breakpoint-containing single reads that align in split fashion to discontiguous loci in the reference genome (split-read). These diverse alignment signals are difficult to integrate and most algorithms use just one. Other methods use two signals, but to our knowledge these limit initial detection to one signal and use the second to add confidence, refine breakpoint intervals, or genotype additional samples [1-3].

An approach that integrates multiple signals allows for more sensitive SV discovery than methods that examine merely one signal, especially when considering heterogeneous samples and/or low coverage data, because each individual read generally produces only one signal type (e.g., read-pair or split-read, but not both). The impact of improved sensitivity is particularly acute in low coverage datasets or in studies of heterogeneous cancer samples where any given variant may only be present in a subset of cells. However, even with high coverage data, integration of multiple signals can increase specificity by allowing for more stringent criteria for reporting a variant call.

# Results

Here, we present a novel and general probabilistic SV discovery framework that naturally integrates multiple SV detection signals, including those generated from read alignments or prior evidence, and that can readily adapt to any additional source of evidence that may become available with future technological advances.

**Overview of LUMPY**

Our framework is based upon a general probabilistic representation of an SV breakpoint that allows any number of alignment signals to be integrated into a single discovery process (Methods). We define a breakpoint as a pair of bases that are adjacent in an experimentally sequenced "sample" genome but not in the reference genome. To account for the varying level of genomic resolution inherent to different types of alignment evidence, we represent a breakpoint with a pair of probability distributions spanning the predicted breakpoint regions (Figure 1, Methods). The probability distributions reflect the relative uncertainty that a given position in the reference genome represents one end of the breakpoint.

Our framework provides distinct modules that map signals from each alignment evidence type to the common probability interval pair. For example, paired-end sequence alignments are projected to a pair of intervals upstream or downstream (depending on orientation) of the mapped reads (Figure 1a). The size of the intervals and the probability at each position is based on the empirical size distribution of the sample's DNA fragment library. The distinct advantage of this approach is that any type of evidence can be considered, as long there exists a direct mapping from the SV signal to a breakpoint probability distribution. Here, we provide three modules for converting SV signals to probability distributions: read-pair, split-read, and generic. However, we emphasize that our framework is readily extensible to new signals that may potentially result from new DNA sequencing technologies or alternative SV detection approaches. The read-pair module maps the output of a paired-end sequence alignment algorithm such as NOVOALIGN (C. Hercus, unpublished [4]) or BWA [5], the split-read module maps the output of a split-read sequence alignment algorithm (e.g., YAHA [6], BWA-SW [7], or BWA-MEM [3], and the generic module allows users to include SV signal types that do not have a specific module implemented (e.g., a priori knowledge such as known SV, and/or output from copy number variation discovery tools).

Once the evidence from the different alignment signals is mapped to breakpoint intervals, overlapping intervals are clustered and the probabilities are integrated (see Methods for details). Any clustered breakpoint region that contains sufficient evidence (based on user-defined arguments) is returned as a predicted SV. The resolution of the predicted breakpoint regions is improved by trimming the positions with probabilities in the lower percentile of the distribution (e.g., the lowest 0.1 percent).

It is well established that variant calling is improved by integrating data from multiple samples [3, 8-10]. The LUMPY framework naturally handles multiple samples by tracking the sample origin of each probability distribution during clustering (Figure 1b, Methods). As an example of a typical analysis, LUMPY can identify SVs in a whole-genome, 50X coverage paired-end Illumina dataset from the NA12878 CEPH individual in 12.2 hours using 8 Gb of memory using a single processor (Methods).

Given that these performance characteristics are comparable to existing approaches for SNP and INDEL detection, and that there is an approximately linear relationship between data volume, time and memory usage, we anticipate that simultaneous analysis of tens and eventually hundreds of human genomes will be possible with LUMPY using commodity hardware.

We implemented LUMPY in an open source C++ software package (available at https://github.com/arq5x/lumpy-sv) that is capable of detecting structural variation from multiple alignment signals in BAM [11] files from one or more samples.

**Performance comparisons**

We compared LUMPY's performance to three other popular and actively maintained SV discovery packages: GASVPro [2] , DELLY [1] and PINDEL [12]. These algorithms were selected due to their widespread use and inclusion in large-scale projects such as The Cancer Genome Atlas (GASVPro) and 1000 Genomes (DELLY and PINDEL). Moreover, PINDEL was one of the first published SV discovery tools, and GASVPro and DELLY both consider a secondary SV signal along with paired-end alignments (read-depth and split-read, respectively). Both GASVPro and DELLY have also demonstrated substantial improvement over other popular SV tools such as Breakdancer [13] and HYDRA [14].

Detection performance was measured using both simulated data and previously published Illumina sequencing data from the widely studied NA12878 CEPH individual. The first simulation measured each tool's basic detection capabilities in a prototypical scenario by simulating 2500 homozygous variants from various SV classes at random genomic locations. The second simulation assessed the power of each tool to detect 5516 known deletion variants present at varying allele frequencies within a mixed sample, as often occurs in heterogeneous tumors. Lastly, the analysis of SVs in the NA12878 genome assessed the performance of each tool on real data containing sequencing errors and other detection confounders that are difficult to simulate. In addition, the analysis of NA12878 (and her parents) measured the effect of considering either multiple samples or prior SV knowledge on LUMPY's performance. In each case, we measured performance in terms of sensitivity and false discovery rate (FDR) by comparing the predicted SV breakpoints to known breakpoints. Predictions that overlapped simulated variants were considered true positives, and all other predictions were considered false positives (see Methods).

The simulated results were base on alignments generated by NOVOALIGN, and the NA12878 results were based on BWA alignments. In both cases, YAHA was used to generate split-read alignments. LUMPY was also tested using alternative read alignment pipelines using either BWA-backtrack or BWA-MEM for paired-end alignment, and BWA-MEM for split-read alignment (with roughly similar results; see Figure 2d and 2e).

**Homozygous variants of different SV types**

To assess the impact of coverage, SV type and SV size on algorithm performance, we first simulated a set of sample genomes that included 2500 deletions, tandem duplications, inversions and translocations, randomly placed throughout the human genome (build 37). Variants were created with an in-house SV simulation tool (SVSIM) that creates defined alterations to the reference genome (Faust and Hall, unpublished). For each SV type, the variant size ranged from 100bp to 10kb. We then used the WGSIM read simulator (H. Li, unpublished [15]) to "sequence" each simulated genome at 2X, 5X, 10X, 20X, and 50X haploid coverage.

LUMPY was consistently more sensitive than the other algorithms across nearly all coverage levels and SV types (Figure 2a). DELLY detects three more translocations than LUMPY at 20X coverage, at the expense of 93 more false positives. LUMPY and DELLY were the only algorithms that detected all variant types; GASVPro and PINDEL do not support detection of tandem duplications (tandem duplication support has since been added). LUMPY's superior sensitivity was most dramatic in lower coverage tests (<10X). For example, LUMPY detected 32.4% and 87.2% of all deletions at 2X and 5X coverage, respectively, whereas GASVPro detected 7.4% and 49.8%, DELLY detected 8.4% and 57.2%, and PINDEL detected 7.4% and 50%. At best, LUMPY was 35.5 times more sensitive than PINDEL for detecting translocations at 5X coverage (69.1% vs. 2%). At worst, LUMPY was 1.06 times more sensitive that DELLY for detecting translocations at 2X coverage (69.1% vs. 65.1%). At higher coverage (10-50X), LUMPY's sensitivity advantage persisted; it ranged from 88.8% to 99.6% across all SV types, whereas GASVPro ranged from 14.3% to 94.3%; DELLY ranged from 76.7% to 96.8%; and PINDEL ranged from 0.2% to 96.5% (excluding the SV types which GASVPro and PINDEL are incapable of detecting).

LUMPY's FDR remained low (less than 2%) in all but the highest coverage cases (Figure 2b), and there was only one instance where LUMPY's FDR that was more than 2 percentage points higher than the best performing tool (GASVPro's FDR for inversions at 50X was 1.6% while LUMPY's was 4.5%). In general, the FDR for LUMPY, GASVPro, and PINDEL increased as coverage increased, ranging from 0% to 7.2% fro LUMPY, 0% to 7.1% for GASVPro, and from 0% to 53.2% for PINDEL. In contrast, the FDR for DELLY dropped as coverage went from 2X to 20X, then increased again at 50X, ranging between 3.3% and 35.2%. These patterns suggest that coverage-based parameter tuning could be used to minimize FDR for all the tools. We also note that FDR calculations depend on the number of true positives, which vary widely across SV varieties (Figure 2c). In certain cases (e.g., translocations), LUMPY has a far higher absolute number of false positives, but these are counterbalanced by a much higher number of true positives as well. Alternatively, in cases where a specific SV type is not supported and no true positive calls were made (i.e., GASVPro and PINDEL for tandem), the FDR can reach 100%.

**Heterogeneous tumor simulation**

Variant detection is especially challenging in tumor studies because biopsied samples generally include a mixture of abnormal and normal tissue, and because many tumor samples are composed of multiple clonal lineages defined by distinct somatic mutations. To assess the performance of our algorithm in this more realistic scenario where increased sensitivity is crucial, we simulated heterogeneous samples by pooling reads from an "abnormal" genome and a "normal" genome at varying ratios. The source of the simulated *abnormal* genome was the human reference genome (build 37) modified (using SVSIM) with 5516 non-overlapping deletions identified by the 1000 Genomes Project [16], and an unmodified human reference genome was used to simulate the *normal* genome. As above, each genome was "sequenced" using WGSIM, and the reads from the two genomes were combined in varying proportions to create a single heterogeneous sample. The ratio of the reads from the abnormal genome (SV allele frequency) varied between 5% and 50%, and the total coverage ranged from 10X and 80X. For example, to simulate a sample with a 5% SV allele frequency at 10X coverage, the abnormal genome was sequenced at 0.5X coverage and the normal genome at 9.5X coverage: when combined, the two sets of reads represent a single heterogeneous sample sequenced at 10X coverage.

LUMPY was more sensitive than GASVPro, DELLY, and PINDEL in all cases, especially when the coverage of the abnormal genome was low owing to either lower coverage, low SV allele frequency, or both (Figure 3a). For example, at 10X coverage and 20% SV allele frequency LUMPY detects 31% of the SVs, whereas GASVPro, DELLY, and PINDEL detect only 9.2%, 10.9%, and 6.2% of the SVs, respectively. This represents a 2.9-fold increase in sensitivity over the next best method. In general, to achieve the same level of sensitivity, GASVPro, DELLY and PINDEL required roughly twice as much evidence as LUMPY (by either increased coverage or SV allele frequency). For example, at 20X coverage LUMPY detected 6.2% of variants present at 5% SV allele frequency, whereas GASVPro, DELLY, and PINDEL required 10% SV allele frequency to achieve similar sensitivity (10.3%, 8.8%, and 5.5%, respectively). We note that this trend is also apparent across SV varieties in the previous homozygous test (Figure 2a).

The FDR for LUMPY was lower than all other tools in all cases (Figure 3b), ranging from 0% to 3.4%. For GASVPro, DELLY, and PINDEL, the FDR is particularly high when SV allele frequency is low. For example, at 10X coverage the FDR for GASV is 5.8 times higher at 5% SV allele frequency than at 50% frequency, DELLY FDR is 3.5 times higher, and PINDEL FDR is 111.9 times higher. At 20X coverage these differences are 4.1, 10.9, and 21.5 times higher at 5% frequency than at 50% frequency for GASVPro, DELLY, and PINDEL, respectively. This is in contrast to LUMPY, where modest coverage-associated increases to FDR can likely be managed via parameter tuning, without significantly decreasing sensitivity.

**SV detection in the NA12878 genome**

Although it is difficult to precisely measure the sensitivity and accuracy of SV predictions made from a real data set, it is also important to evaluate each tool's performance when confronted with real data containing artifacts that are not easily captured by simulations (e.g., PCR artifacts, chimeric molecules, reads from poorly assembled genomic regions, etc.). In this experiment we compared SV detection performance in the NA12878 individual by analyzing the Illumina Platinum Genomes dataset, which represents ~50X coverage of the NA12878 genome (European Nucleotide Archives; ERA172924). We additionally subsampled this dataset to ~5x coverage to assess SV detection in low coverage scenarios.

To estimate sensitivity and FDR, we compared predictions made by each tool to two truth sets: the first set was based on 3376 validated, non-overlapping deletions from the 1000 Genomes project [16], and the second set included the same 3376 validated deletions as well as 2128 additional deletion predictions made by at least two tools using the 50X dataset (5504 total deletions). The rationale for two truth sets is that, although the 1000 Genomes callset is to our knowledge the most comprehensive set of deletions for NA12878, it still represents only a subset of the actual deletions in that individual's genome. In this study we have the benefit of higher quality sequencing data, longer reads and improved SV detection tools, and thus are likely to discover novel deletions that were missed by Mills et al. [16]. Furthermore, since PINDEL was one of the tools used to generate the 1000 Genomes callset [16], it is biased against predictions made by LUMPY, GASVPro and DELLY. Therefore, although we acknowledge that some of the predictions included in the larger truth set may not reflect genuine SVs, these calls nonetheless represent real differences between the sequencing data and the reference genome and, as such, provide a more realistic estimate of the FDR for each tool.

A unique strength of LUMPY relative to other tools is the ability to consider different types of evidence from multiple samples. To demonstrate this capability, we included results for three different LUMPY configurations: (1) the standard configuration of read-pair and split-read signals from

NA12878; (2) read-pair and split-read signals from NA12878, as well as prior knowledge (using LUMPY's generic evidence module) of deletions discovered by the 1000 Genomes Project using low coverage whole genome sequencing (phase 1, release version 3); (3) read-pair and split-read signals from both NA12878 and her parents (NA12891 and NA12892). The last two calling strategies are specific to LUMPY and are intended to demonstrate both the ability and the benefit of including data from different samples and from prior results.

At 5X coverage (Figure 4a), LUMPY was more sensitive than both GASVPro and PINDEL (17.6% versus 9.5% and 16.1% for the first truth set and 13.5% versus 7.6% and 10.8% for the second truth set) and had an either equivalent or better FDR (33.7% versus 56.1% and 33.4% for the first truth set and 18.9% versus 43.9% and 28.5% for the second). While DELLY was more sensitive than LUMPY (19.3% for the first truth set and 16.1% for the second), it was at the expense of at least 0.5-fold higher FDR (51.5% and 35.1%). However, when LUMPY is provided with priors from the 1000 Genomes low coverage deletion calls, sensitivity is increased to 22.5% with a negligible effect on false positives (leading to a lower FDR). Sensitivity is further improved to 27.0% when LUMPY performs simultaneous variant calling on NA1287 and her parents, with a similarly small effect on FDR, which clearly demonstrates the benefit of pooled variant calling on genetically related samples. Overall, the effect of the second truth set at lower coverage was similar between LUMPY, GASVPro, and DELLY with FDR decreasing by 14.8, 12.2, and 16.4 percentage points, respectively. The effect on PINDEL was the smallest, with FDR decreasing by only ~5 percentage points; however, we note that this smaller effect is expected considered that PINDEL was used in part to create the baseline truth set [16].

At 50X coverage (Figure 4b) and considering the first truth set, LUMPY, PINDEL and DELLY had similar sensitivity (62.1%, 63% and 61.3%, respectively), and all outperformed GASVPro (55.1%). Owing to the incompleteness of the first truth set, all of the tools had high FDR. LUMPY performed the best at 54%, followed by GASVPro at 70.9%, PINDEL at 72%, and DELLY at 84.8%. When considering the second truth set, the effect on performance was dramatic for all the tools except PINDEL (as expected). LUMPY had the highest sensitivity at 73.3%, followed by DELLY at 71.3%, GASVPro at 61.9%, and PINDEL where the sensitivity actually dropped to 51.8% (due to the increased size of the second truth set). LUMPY also had the lowest FDR at 14.8%, followed by GASVPro at 42%, PINDEL at 63.1%, and DELLY at 71.5%. As expected given the substantially higher coverage, the inclusion of either priors from known SVs or the parental genomes had little effect on LUMPY's sensitivity. Overall, these results demonstrate that LUMPY provides substantial improvements in discovery sensitivity over existing methods while also maintaining equivalent or lower false discovery rates.

Importantly, while the comparisons presented in Figure 4 are based upon the choice of a single detection threshold chosen for each tool (see Methods), LUMPY's detection accuracy remains superior across a broad spectrum of thresholds (Figure 5), indicating that the framework itself – not arbitrary parameter choices – underlie LUMPY's increased performance.

**Benefits of integrating all signals for SV discovery**

LUMPY's superior sensitivity in these performance tests is a direct consequence of the fact that it simultaneously integrates multiple SV detection signals during SV discovery. The benefits of this approach are clear from the super-additive effect of combining read-pair and split-read signals within the LUMPY framework, relative to using either signal alone (Figure 3c). In contrast, although other tools such as GASVPro [2], DELLY [1] and Genome STRiP [3] also exploit multiple SV detection signals, to our knowledge they first use one signal (i.e., read-pair) to drive discovery and then refine and/or genotype candidates with a second signal (i.e., split-read or read-depth). An intrinsic limitation of stepwise integration is that other available signals cannot increase the number of true positive SV calls beyond those candidates identified by the signal used for initial discovery. Consistent with this interpretation, inclusion of a second SV detection signal has little to no effect on DELLY's or GASVPro's sensitivity (Figure 3c) relative to using the primary read-pair signal alone.

## Discussion

We have developed a general probabilistic framework for SV discovery, and have demonstrated that our framework is more sensitive than existing discovery tools across all SV types and coverage levels, and in both real and simulated human genome datasets. Importantly, LUMPY's increased sensitivity does not come at the cost of excessive spurious SV predictions. LUMPY therefore represents an important technological advance, especially in the context of cancer genomics where sensitivity is crucial for identifying low frequency variants within heterogeneous tumor samples.

LUMPY's high sensitivity is a direct consequence of combining multiple SV detection signals. LUMPY integrates disparate signals by converting them to a common format in which the two predicted breakpoint intervals in the reference genome are represented as paired probability distributions. SV prediction operations are then performed at this higher level. This novel approach has the key advantage that any SV detection signal can be integrated into the framework so long as a breakpoint probability can be assigned to each base pair in a candidate breakpoint region. Potential detection signals include paired-end and split-read alignments, alignments from assembled contigs, raw read-depth measurements, or CNVs detected by segmentation of read-depth, array comparative genomic hybridization or SNP array data. As we demonstrate using the NA12878 genome, inclusion of previously discovered SVs as priors can significantly enhance SV discovery sensitivity, which is an example of the flexibility and generality of our framework. To facilitate integration, each evidence type can be assigned a different weight reflecting the user's prior expectations. As sequencing technologies and SV detection strategies evolve, new sources of evidence can be easily incorporated without modifying the underlying logic of the SV detection algorithm itself; the sole requirement is the development of a new module that maps the SV detection signal to a paired probability distribution.

In addition to facilitating signal integration, our use of probability distributions also enables more accurate prediction of breakpoint position, since in most cases not all coordinates within a predicted breakpoint interval are equally likely to be the breakpoint. By using probability distributions, this spatial uncertainty is propagated throughout the SV detection process. Following SV detection, LUMPY can report the final integrated probability distribution for each predicted variant to allow for comparison across studies. Alternatively, the final breakpoint probability distributions from one study could be used as a source of prior evidence in another.

How could LUMPY's performance be further improved? First and foremost, inclusion of read-depth information should significantly improve performance at duplication and deletion variants. At present, this can be accomplished by converting the output of copy number segmentation tools to breakpoint probability distributions, and providing these to LUMPY using the generic module (Figure 1a); however, we expect that more significant improvements will be possible using raw read-depth data. Second, for applications that require ultra-sensitive detection of known structural variants – such as low coverage population scale sequencing – LUMPY could be packaged with dataset priors reflecting the positions and allele frequencies of previously identified SVs. While we show that this is feasible using the existing LUMPY framework (Figure 4), we note that substantial improvements to sensitivity may require more comprehensive and accurate SV catalogs than are currently available. Finally, a major challenge for SV detection is distinguishing bona fide variants from false positives caused by alignment artifacts and other sources of error. In this respect, breakpoint probability distributions provide a highly quantitative source of information regarding the relative spacing of discordant and/or split alignments at a locus. By training on a set of known variants, it should be possible to derive a probabilistic measure of variant confidence that is based not only on the number of clustered reads, but also on the shape of the final integrated probability distribution. Alternatively, knowledge of the shape of "true" breakpoint probability distributions could potentially be used as an objective function during read clustering.

In a more general sense, our approach for integrating SV detection signals - in essence, performing genome interval comparison operations using probability distributions rather than "flat" features - could be useful for any application that involves comparison of genomic features whose exact coordinates are unknown, and whose positional uncertainty can be represented rationally in the form of a probability distribution. Rapid and efficient probabilistic comparisons could be enabled through extensions to existing interval-based software such as BEDTools [17, 18].

**Materials and Methods**

We propose a breakpoint prediction framework that can accommodate multiple classes of evidence from multiple sources in the same analysis. Our framework makes use of an abstract breakpoint evidence type to define a set of functions that serve as an interface between specific evidence subtypes (e.g., paired-end sequence alignments and split-read mappings) and the breakpoint type. Any class of evidence for which these functions can be defined may be included in our framework. To demonstrate the applicability of this abstraction, we defined three breakpoint evidence subtypes: read-pair, split-read, and a general breakpoint interface.

Since our framework combines evidence from multiple classes, it extends naturally to include evidence from multiple sources. The sources that can be considered in a single analysis may be any combination of evidence from different samples, different evidence subclasses from a single sample, or prior information about known variant positions. We refer to a given set of data as a breakpoint evidence instance, and assume that each instance contains only one evidence subtype and is from a single sample. To help organize the results of analysis with multiple samples or multiple instances for a single sample, each instance is assigned an identifier that can be shared across instances.

**Breakpoint**

A breakpoint is a pair of genomic coordinates that are adjacent in a sample genome but not in a reference genome. Breakpoints can be detected, and their locations predicted by various evidence classes such as paired-end sequence alignments or split-read mappings. To support the inclusion of different evidence classes into a single analysis, we define a high-level breakpoint type as a collection of the evidence that corroborates the location and variety (e.g., deletion, tandem duplication, etc.) of a particular breakpoint. Since many evidence classes provide a range of possible breakpoint locations, we represent the breakpoint's location with a pair of breakpoint intervals where each interval has a start position, an end position, and a probability vector that represents the relative probability that a given position in the interval is one end of the breakpoint. More formally, a breakpoint is a tuple $b = \langle E,l,r,v \rangle$, where $b.E$ is the set of evidence that corroborates the location and variety of a particular breakpoint; $b.l$ and $b.r$ are left and right breakpoint intervals each with values $b.l.s$ and $b.l.e$ that are the start and end genomic coordinates and $b.lp$ is a probability vector where $|b.l.p| = b.l.e - b.l.s$ and $b.l.p[i]$ is the relative probability that position $b.l.s + i$ is one end of the breakpoint (similar for $b.r$); and $b.v$ is the breakpoint variety. Within the context of this method, breakpoint variety determinations are based on the orientation of the evidence. It is important to note that while a breakpoint may be labeled as a deletion when it contains evidence from a paired-end sequence alignment with a +/- orientation, the breakpoint may in fact be the result of some other event or series of complex events.

If there exist two breakpoints $b$ and $c$ in the set of all breakpoints $B$ where $b$ and $c$ intersect ($b.r$ intersects $c.r$, $b.l$ intersects $c.r$, and $b.v = c.v$), then $b$ and $c$ are merged into interval $m$, $b$ and $c$ are removed from $B$, and $m$ is placed into $B$. The evidence set $m.E$ is the union of the evidence sets $b.E$ and $c.E$.

A straight-forward method to define breakpoint intervals $m.l$ and $m.r$ would be to let $m.l.s = \max(b.l.s, c.l.s)$ and $m.l.e = \min(b.l.e, c.l.e)$, similar for $m.r$. However, if a spurious alignment is merged into a set of genuine breakpoints, the resulting breakpoint interval can be "pulled" away from the actual breakpoint. The impact of an outlier can be minimized or eliminated once the full set of corroborating alignments is collected for a given breakpoint, but collecting the full set is complicated by the fact that alignments are considered in order and outliers typically occur first. To account for this, we define a liberal merge process where $m.l.s$ is the mean start position for the left intervals in $m.E$, and $m.l.e$ is the mean end position for the left intervals in $m.E$, similar for $m.r$.

Once all the evidence has been considered, an SV call $s$ (also a breakpoint) is made for each breakpoint $b \in B$ that meets a user-defined minimum evidence threshold (e.g., four pieces of evidence). The boundaries of the breakpoint intervals $s.l$ and $s.r$ are the trimmed mixture distributions of the left and right intervals in $b.E$. Let $s.l.s = \min(\{e.l.s \mid e \in b.E\})$, $s.l.e = \min(\{e.l.e \mid e \in b.E\})$, and $s.l.p[i] = \sum_{e \in b.E} e.l.p[i-o]$ where $o$ is the offset value $e.l.s - s.l.s$ (similar for $s.r$). The value at $s.l.p[i]$ (or $s.r.p[i]$) represents the level of agreement among the evidence in $b.E$ that position $i$ is one end of the breakpoint. The intervals $s.l$ and $s.r$ are then trimmed to include only those positions that are in the top percentile (e.g., top 99.9 percent of values). An outlier in $b.E$ will extend the interval $s.l$, but the extended region will have little support from other elements in $b.E$ and values of $s.p$ in that region will be relatively small and are likely to be removed in the trimming process.

**Breakpoint Evidence**

To combine distinct SV alignment signals such as read-pair and split-read alignments with the general breakpoint type defined above, we define an abstract breakpoint evidence type. This abstract type defines an interface that allows for the inclusion of any data that can provide the following functions: `is_bp` determines if a particular instance of the data contains evidence of a breakpoint, `get_v` determines the breakpoint variety (e.g., deletion, tandem duplication, inversion, etc.), and `get_bpi` maps the data to a pair of breakpoint intervals.

To demonstrate the applicability of this abstraction, we defined three breakpoint evidence instances: paired-end sequencing alignments, split-read alignments, and a general breakpoint interface. Read-pair and split-read are among the most frequently used evidence types for breakpoint detection, and the general interface provides a mechanism to include any other sources of information such as known variant positions or output from other analysis pipelines (e.g., read-depth calls). As technologies evolve and our understanding of structural variation improves, other instances can be easily added.

**Paired-End Alignments**

Paired-end sequencing involves fragmenting genomic DNA into roughly uniformly-sized fragments, and sequencing both ends of each fragment to produce paired-end reads $\langle x, y \rangle$, which we will refer to as "read-pairs". Each read is aligned to a reference genome $R(x) = \langle c, o, s, e \rangle$, where $R(x).c$ is the chromosome that $x$ aligned to in the reference genome, $R(x).o = +/-$ indicates the alignment orientation, and $R(x).s$ and $R(x).e$ delineate the start and end positions of the matching sequence within the chromosome. We assume that both $x$ and $y$ align uniquely to the reference and that $R(x).s < R(x).e < R(y).s < R(y).e$. While in practice it is not possible to know the position of read $x$ in the sample genome (in the absence of whole-genome assembly), it is useful to refer to $S(x) = \langle o, s, e \rangle$ as the alignment of $x$ with respect to the originating sample's genome.

Assuming that genome sequencing was performed with the Illumina platform, read-pairs are expected to align to the reference genome with a $R(x).o = +$, $R(y).o = -$ orientation, and at distance $R(y).e - R(x).s$ roughly equivalent to the fragment size from the sample preparation step. Any read-pair that aligns with an unexpected configuration can be evidence of a breakpoint. These unexpected configurations include matching orientation $R(x).o = R(y).o$, alignments with switched orientation $R(x).o = -$, $R(y).o = +$, and an apparent fragment length $(R(y).e - R(x).s)$ that is either shorter or longer than expected. We estimated the expected fragment length to be the sample mean fragment length $l$, and the fragment length standard deviation to be the sample standard deviation $s$ from the set of properly mapped read-pairs (as defined by the SAM specification) in the sample data set. Considering the variability in the sequencing process, we extend the expected fragment length to include sizes $l + v_l s$, where $v_l$ is a tuning parameter that reflects spread in the data.

When $x$ and $y$ align to the same chromosome ($R(x).c = R(y).c$), the breakpoint variety can be inferred from the orientation of $R(x)$ and $R(y)$. If the orientations match, then the breakpoint is labeled as an inversion, and if $R(x).o = -$ and $R(y).o = +$ then the breakpoint is labeled as a tandem duplication. Any breakpoint with the orientation $R(x).o = +$ and $R(y).o = -$ is labeled as a deletion. When $x$ and $y$ align to different chromosomes ($R(x).c \neq R(y).c$), the variety is labeled inter-chromosomal. At present, LUMPY does not explicitly support identification of insertions that are spanned by paired-end reads, however, if desired these can be identified in a post-processing step through assessment of "deletion" calls.

To map $\langle x, y \rangle$ to breakpoint intervals $l$ and $r$, the ranges of possible breakpoint locations must be determined and probabilities assigned to each position in those ranges. By convention, $x$ maps to $l$ and $y$ to $r$, and for the sake of brevity we will focus on $x$ and $l$ since the same process applies to $y$ and $r$. Assuming that a single breakpoint exists between $x$ and $y$, then the orientation of $x$ determines if $l$ will be upstream or downstream of $x$. If the $R(x).s = +$, then the breakpoint interval begins after $R(x).e$ (downstream), otherwise the interval ends before $R(x).s$ (upstream).

The length of each breakpoint interval is proportional to the expected fragment length $L$ and standard deviation $s$. Since we assume that only one breakpoint exists between $x$ and $y$, and that it is unlikely that the distance between the ends of a pair in the sample genome ($S(y).e - S(x).s$) is greater than $L$, then it is also unlikely that one end of the breakpoint is at a position greater than $R(x).s + L$, assuming that $R(x).o = +$. If $R(x).o = -$, then it is unlikely that a breakpoint is at a position less than $R(x).e - L$. To account for variability in the fragmentation process, we extend the breakpoint to $R(x).e + (L + v_f s)$ when $R(x).o = +$, and $R(x).s - (L + v_f s)$ when $R(x).o = -$, where $v_f$ is a tuning parameter that, like $v_l$, reflects the spread in the data.

The probability that a particular position $i$ in the breakpoint interval $l$ is part of the actual breakpoint can be estimated by the probability that $x$ and $y$ span that position in the sample. For $x$ and $y$ to span $i$, the fragment that produced $\langle x, y \rangle$ must be longer than the distance from the start of $x$ to $i$, otherwise $y$ would occur before $i$ and $x$ and $y$ would not span $i$ (contradiction). The resulting probability is $P(S(y).e - S(x).s > i - R(x).s)$ if $R(x).o = +$, and $P(S(y).e - S(x).s > R(x).e - i)$ if $R(x).o = -$. While we cannot directly measure the sample fragment length ($S(y).e - S(x).s$), we can estimate its distribution by constructing a frequency-based cumulative distribution $D$ of fragment lengths from the same sample that was used to find $L$ and $s$, where $D(j)$ gives the proportion of the sample with fragment length greater than $j$.

**Split-Read Alignments**

A split-read alignment is a single DNA fragment $X$ that does not contiguously align to the reference genome. Instead, $X$ contains a set of two or more substrings $x_i...x_j$ ($X = x_1x_2...x_n$), where each substring aligns to the reference $R(x_i) = \langle c, o, s, e \rangle$, and adjacent substrings align to non-adjacent locations in the reference genome $R(x_i).e \neq R(x_i + 1).s + 1$ or $R(x_i).c \neq R(x_i + 1).c$ for $1 \leq i \leq n - 1$. A single split-read alignment maps to a set of adjacent split-read sequence pairs ($\langle x_1, x_2 \rangle, \langle x_2, x_3 \rangle, ..., \langle x_{n-1}, x_n \rangle$), and each pair $\langle x_i, x_i+1 \rangle$ is considered individually.

By definition, a split-read mapping is evidence of a breakpoint and therefore the function `is_bp` trivially returns true.

Both orientation and mapping location must be considered to infer the breakpoint variety for $\langle x_i, x_i+1 \rangle$. When the orientations match $R(x_i).o = R(x_i + 1).o$, the event is marked as either a deletion or a tandem duplication. Assuming that $R(x_i).o = R(x_i + 1).o = +$, $R(x_i).s < R(x_i + 1).s$ indicates a gap caused by a deletion and $R(x_i).s > R(x_i + 1).s$ indicates a tandem duplication. These observations are flipped when orientations $R(x_i).o = R(x_i + 1).o = -$. When the orientations do not match $R(x_i).o \neq R(x_i + 1).o$, the event is marked as an inversion and the mapping locations do not need to be considered. When $x$ and $y$ align to different chromosomes, the variety is marked as inter-chromosomal. LUMPY does not currently attempt to identify insertions that are completely contained within a long read, but this will be supported in future versions. We note that this capability requires long-read aligners to report the number and order of alignments within a read (which is not formally supported in the current SAM format specifications).

The possibility of errors in the sequencing and alignment processes creates some ambiguity in the exact location of the breakpoint associated with a split-read alignment. To account for this, each alignment pair $\langle x_i, x_i + 1 \rangle$ maps to two breakpoint intervals $l$ and $r$ centered at the split. The probability vectors $l.p$ and $r.p$ are highest at the midpoint and decrease exponentially toward their edges. The size of this interval is a configurable parameter $v_s$ and is based on the quality of the sample under consideration and the specificity of the alignment algorithm used to map the sequences to the reference genome.

Depending on the breakpoint variety, the intervals $l$ and $r$ are centered on either the start or the end of $R(x_i)$ and $R(x_i + 1)$. When the breakpoint is a deletion $l$ is centered at $R(x_i).e$ and $r$ at $R(x_i + 1).s$, and when the breakpoint is a tandem duplication $l$ is centered at $R(x_i).s$ and $r$ at $R(x_i + 1).e$. If the breakpoint is an inversion, $l$ and $r$ are both centered either at the start positions or end positions of $R(x_i)$ and $R(x_i + 1)$, respectively. Assuming that $R(x_i).s < R(x_i + 1).s$, if $R(x_i).o = +$ then $l$ and $r$ are centered at $R(x_i).e$ and $R(x_i + 1).e$, otherwise they are centered at $R(x_i).s$ and $R(x_i + 1).s$. If $R(x_i).s > R(x_i + 1).s$, then the conditions are swapped.

**Generic Evidence**

The generic evidence subclass provides a mechanism to directly encode breakpoint intervals using the BEDPE format [17]. BEDPE is an extension of the popular BED format that provides a means to specify a pair of genomic coordinates; in this case the pair represents the two breakpoint positions in the reference genome. This subclass extends our framework to include SV signal types that do not yet have a specific subclass implemented. For example, the a set of variants that are known to exist in the population can be included in the analysis of an individual or variants that are known to exist in a particular type of cancer can be included in the analysis of a tumor. This signal can be included in the analysis by expanding the edges of the predicted intervals to create breakpoint intervals, and encoding these intervals in BEDPE format. Each BEDPE entry is assumed to be a real breakpoint (`is_bp`), the variety is encoded in the auxiliary fields in BEDPE (`get_v`), and the intervals are directly encoded in BEDPE (`get_bpi`).

**Performance comparisons**

Both simulated and real datasets were used to compare the sensitivity and false discovery rate of LUMPY to other SV detection algorithms (GASVPro, DELLY, and PINDEL). Two types of simulations were performed: one in which homozygous variants of diverse varieties were introduced at random positions throughout the reference genome, and another in which a heterogeneous tumor sample was simulated by mixing reads from a modified "abnormal" human reference genome (containing 1000 Genomes deletions) and a unmodified "normal" human reference genome in varying proportions. We also used publicly available Illumina sequencing data of the NA12878, NA12891, and NA12892 individuals. Two scenarios were considered: the original 50X coverage files, and 5X subsamples of the original data sets.

In the case of the homozygous simulation, we used SVSIM to create new versions of the human reference genome (build 37) containing 2500 simulated variants of each variety. For deletions, tandem duplications and inversions we randomly placed 2500 non-overlapping variants ranging from 100 bp to 10000 bp in size. To simulate translocations, we randomly inserted 2500 non-overlapping inter-chromosomal regions of 1000 bp, derived from random donor sites in the reference genome. Although we note that the true variant variety in this case is actually an insertion, the inserted segment exceeds the insert size of the sequencing library as well as the read length, and thus the breakpoints formed by such insertions accurately simulate a translocation. Each simulated genome was sampled to 40X, 20X, 10X, 5X, and 2X coverage.

To simulate a heterogeneous tumor sample, we combined simulated reads from both a modified and unmodified version of the human reference genome (build 37). The modified genome was created using SVSIM, and included 5516 non-overlapping deletions identified by the 1000 Genomes Project. Each simulation combined reads from both the modified and unmodified genomes in varying proportions. We refer to the proportion of reads that were derived from the modified genome as the SV allele frequency. The simulated SV allele frequencies were 5%, 10%, 20% and 50%, and the simulated coverages were 10X, 20X, 40X, and 80X. For example, in the simulation with 5% SV allele frequency and 10X coverage, the modified genome was sampled at 0.5X coverage and the unmodified genome was sampled at 9.5X coverage. The two sets of reads are then pooled into a single 10X coverage sample.

For all simulations, WGSIM was used to sample paired-end reads with a 150 bp read length, a 500 bp mean outer distance with a 50 bp standard deviation, and default error rate settings. Paired-end reads were mapped to the reference genome with NOVOALIGN version V2.07.08, using the random repeat reporting and allowing only one alignment per read. From the NOVOALIGN output, all soft-clipped (≥20 bp clipped length) and unmapped reads were realigned with the split-read aligner YAHA using a word length of 11 and a minimum match of 15. The NOVOALIGN output was used as input to DELLY ,GASVPro, and PINDEL, and both NOVOALIGN and YAHA output were used as input to LUMPY. In all algorithms, the minimum evidence threshold was four. For LUMPY, the tuning parameters `min_non_overlap` was set to 150, `discordant_z` was set to 4, `back_distance` was set to 20, weight was set to 1, and `min_mapping_threshold` was set to 1. For GASVPro, `LIBRARY_SEPARATED` was set to all, `CUTOFF_LMINLMAX` was set to SD=4, `WRITE_CONCORDANT` was set to true, and `WRITE_LOWQ` was set to true. For DELLY, `map-qual` was set to 1, and the `inc-map` flag was set. DELLY paired-end (pe) and split-read (sr) calls were combined into a single paired-end and split-read (pe+sr) call set by taking the union of the two sets where the split-read call was retained when a call was common to both sets. For PINDEL, `minimum_support_for_event` was set to 4, all chromosomes were considered, and `report_interchromosomal_events` was set to true.

For the real data, LUMPY, GASVPro, DELLY, and PINDEL considered Illumina sequencing of the NA12878 individual. The LUMPY trio results also considered sequencing data from that individual's parents, NA12891 and NA12892. All sequencing samples were retrieved from the European Nucleotide Archives (submission ERA172924), and were previously aligned using BWA. Soft-clipped (≥20 bp clipped length) and unmapped reads were realigned with the split-read aligner YAHA using a word length of 11 and a minimum match of 15. The LUMPY prior results also considered all 1000 Genomes variant calls using the generic evidence module. The original sequencing files were at 50X coverage and were used in the 50X experiments. The 5X experiments considered sequencing files that were created by subsamples 10% of the original paired-end alignments. For all the tools, only deletion predictions on chromosomes 1 though X were considered. Each tool was run with the same options that were used in the simulation experiments, except for the minimum mapping quality for LUMPY, GASVPro, and PINDEL was increased to 10. Since PINDEL uses paired-end reads differently than the other tools, the default mapping quality of 20 was used. Each call required support of at least four. In the LUMPY trio result, a call had to have support of four from at least one individual (NA12878, NA12891, or NA12892) and at least one piece of support from NA12878. The weight for the 1000 Genomes variant calls in the LUMPY prior result was set to 2. For the identification of the truth sets, the Mills et al. study validated 14012 deletions in NA12878 across 11 independent laboratories. Once duplicate predictions were removed, the first truth set contained 3376 non-overlapping deletions. Between the four tools considered here, 4027 predictions were made by at least 2 tools (2035 were made by all four tools), and 2128 of those merged predictions were not observed by Mills et al. The 3376 Mills et al. validated deletions were combined with the 2128 additional predictions to produce a second, more comprehensive, truth set containing 5504 deletion calls.

The SV breakpoints predicted by each algorithm were compared to the known variants. A true positive was a predicted variant call that intersected within 50 bp of both ends of the simulated breakpoints in the reference genome. All other predictions were considered to be false positives. Since the output of DELLY is a single interval, we took the 100 bp regions flanking the ends of the predicted interval as the predicted breakpoint.

## Competing Interests

## Author Contributions

## Acknowledgements

We thank G. Faust for sharing unpublished SV simulation algorithms and for advising simulation experiments, M. Lindberg for code development, A. Ritz and S. Sindi for help with GASVPro, T. Rausch for help with DELLY, and K. Ye for help with PINDEL. This work was supported by an

## References

1. Rausch T, Zichner T, Schlattl A, Stutz AM, Benes V, Korbel JO: **DELLY: structural variant discovery by integrated paired-end and split-read analysis.** *Bioinformatics* 2012, **28:**i333-i339.
2. Sindi SS, Onal S, Peng LC, Wu HT, Raphael BJ: **An integrative probabilistic model for identification of structural variation in sequencing data.** *Genome biology* 2012, **13:**R22.
3. Handsaker RE, Korn JM, Nemesh J, McCarroll SA: **Discovery and genotyping of genome structural polymorphism by sequencing on a population scale.** *Nature Genetics* 2011, **43:**269-276.
4. Hercus C: **Novoalign.** [http://www.novocraft.com].
5. Li H, Durbin R: **Fast and Accurate Short Read Alignment with Burrows-Wheeler Transform.** *Bioinformatics* 2009.
6. Faust GG, Hall IM: **YAHA: fast and flexible long-read alignment with optimal breakpoint detection.** *Bioinformatics* 2012, **28:**2417-2424.
7. Li H, Durbin R: **Fast and accurate long-read alignment with Burrows-Wheeler transform.** *Bioinformatics* 2010, **26:**589-595.
8. DePristo MA, Banks E, Poplin R, Garimella KV, Maguire JR, Hartl C, Philippakis AA, del Angel G, Rivas MA, Hanna M, et al: **A framework for variation discovery and genotyping using next-generation DNA sequencing data.** *Nature Genetics* 2011, **43:**491-498.
9. Hormozdiari F, Hajirasouliha I, McPherson A, Eichler EE, Sahinalp SC: **Simultaneous structural variation discovery among multiple paired-end sequenced genomes.** *Genome Research* 2011, **21:**2203-2212.
10. Quinlan AR, Boland MJ, Leibowitz ML, Shumilina S, Pehrson SM, Baldwin KK, Hall IM: **Genome sequencing of mouse induced pluripotent stem cells reveals retroelement stability and infrequent DNA rearrangement during reprogramming.** *Cell Stem Cell* 2011, **9:**366-373.
11. Li H, Handsaker B, Wysoker A, Fennell T, Ruan J, Homer N, Marth G, Abecasis G, Durbin R: **The Sequence Alignment/Map format and SAMtools.** *Bioinformatics* 2009, **25:**2078-2079.
12. Ye K, Schulz MH, Long Q, Apweiler R, Ning Z: **Pindel: a pattern growth approach to detect break points of large deletions and medium sized insertions from paired-end short reads.** *Bioinformatics* 2009, **25:**2865-2871.
13. Chen K, Wallis JW, McLellan MD, Larson DE, Kalicki JM, Pohl CS, McGrath SD, Wendl MC, Zhang Q, Locke DP, et al: **BreakDancer: an algorithm for high-resolution mapping of genomic structural variation.** *Nature Methods* 2009, **6:**677-681.
14. Quinlan AR, Clark RA, Sokolova S, Leibowitz ML, Zhang Y, Hurles ME, Mell JC, Hall IM: **Genome-wide mapping and assembly of structural variant breakpoints in the mouse genome.** *Genome Research* 2010, **20:**623-635.
15. Li H: **WGSIM.** [https://github.com/lh3/wgsim].
16. Mills RE, Walter K, Stewart C, Handsaker RE, Chen K, Alkan C, Abyzov A, Yoon SC, Ye K, Cheetham RK, et al: **Mapping copy number variation by population-scale genome sequencing.** *Nature* 2011, **470:**59-65.
17. Quinlan AR, Hall IM: **BEDTools: a flexible suite of utilities for comparing genomic features.** *Bioinformatics* 2010, **26:**841-842.
18. Layer RM, Skadron K, Robins G, Hall IM, Quinlan AR: **Binary Interval Search: a scalable algorithm for counting interval intersections.** *Bioinformatics* 2013, **29:**1-7.

## Figure Legends

**Figure 1. LUMPY workflows. (a)** A LUMPY workflow using three different signals (read-pair, split-read and read-depth) from one sample, as well as prior knowledge regarding known variant sites. **(b)** A LUMPY workflow using a single signal type (in this case, read-pair) from three different samples. Note that multi-signal and multi-sample workflows are not mutually exclusive.

**Figure 2. Performance comparison using homozygous variants of various SV types.** To measure sensitivity and accuracy across diverse variant types, 2500 deletions, tandem duplications, inversions or translocations were randomly embedded in the human reference genome prior to read simulation at different levels of sequence coverage (shown beneath each plot). LUMPY and DELLY considered paired-end (*pe*) and split-read (*sr*) alignments, GASVPro considered paired-end alignments and read-depth (*rd*), and PINDEL considered split-read (*sr*) alignments. **(a)** SV detection sensitivity for each tool. LUMPY was the most sensitive in most cases, and had a marked improvement at lower coverage levels. DELLY detected three more translocations than LUMPY at 20X coverage, at the expense of 93 more false positives. **(b)** The false discovery rate (FDR) for each tool. The FDR for LUMPY was low in all but the highest coverage cases, and there was only one instance where LUMPY's FDR that was more than 2 percentage points higher than the next best performing tool. GASVPro and PINDEL do not support tandem duplications, but false calls were made in some cases which resulted in a 100% FDR. **(c)** The absolute number of false positive calls. LUMPY has a high number of false positives in some cases, but these are counterbalanced by a higher number of true positives (part **a**), resulting in a similar FDR (part **b**). To determine the impact that sequence alignment strategies had on SV detection accuracy, LUMPY's sensitivity (**d**) and false discovery rate (**e**) are shown when predicting deletions via different sequence alignment strategies for the homozygous SV simulation presented in (a), (b), and (c). This comparison is based upon the alignment of the same simulated sequences representing an average of 5X coverage for the simulated genome (i.e., the NOVOALIGN + YAHA results in the lightest blue in this figure are identical to the LUMPY (pe+sr) results for 5X coverage). Since BWA-MEM produces both paired-end (pe) and split-read (sr) alignment signals in a single alignment step, it serves as a basis of comparison to the NOVOALIGN (pe) and YAHA (sr) strategy that was used for the presented LUMPY results. BWA-MEM provides better sensitivity than NOVOALIGN when using the paired-end signal alone, yet YAHA provides better sensitivity than BWA-MEM when using the split-read signal alone. However, the sensitivity is *equivalent* with either the BWA-MEM or NOVOALIGN/YAHA strategies when LUMPY integrates *both* alignment signals. Moreover, the false discovery rate for either strategy is extremely low. These results suggest that either alignment strategy is suitable and demonstrates the generality of the LUMPY framework for SV detection.

**Figure 3. Performance comparison using known deletions at varying intra-sample allele frequencies.** To measure SV detection sensitivity and accuracy in the case of a heterogeneous tumor sample, 5516 non-overlapping deletions identified by the 1000 Genomes Project were embedded in an "abnormal" genome prior to read simulation. Simulated reads from the abnormal genome and the unaltered reference genome were mixed in varying proportions to obtain simulated datasets with different SV allele frequencies. Sequencing coverage levels are shown above each plot, and SV allele frequencies are shown beneath each plot. **(a)** SV detection sensitivity for the four tools. LUMPY and DELLY considered paired-end (*pe*) and split-read (*sr*) alignments, GASVPro considered paired-end alignments and read-depth (*rd*), and PINDEL considered split-read (*sr*) alignments. In all cases, LUMPY was more sensitive than GASVPro, DELLY, and PINDEL, and showed a marked improvement when the coverage of the abnormal genome was low due either to low sequence coverage or low SV allele frequency. In general, to achieve the same level of sensitivity as LUMPY, the other tools required twice as much evidence from the abnormal genome. **(b)** The false discovery rate (FDR) for each tool. The FDR for LUMPY was better than all other tools in all cases, with a notable improvement when SV allele frequency was low. **(c)** The change in sensitivity when considering two SV detection signals versus a single signal alone is shown for the three tools at 40x coverage and at different SV allele frequencies. At low SV allele frequencies (e.g., 5%), LUMPY's use of two signals (i.e., *pe+sr*) has a super-additive effect on sensitivity relative to either signal alone (i.e., *pe* or *sr*), whereas the sensitivity of GASVPro and DELLY was either unchanged or modestly improved with one signal versus two.

**Figure 4. Performance comparison of deletion detection in high and low coverage Illumina sequencing data from the NA12878 individual.** To measure each tool's performance on data that is representative of a typical experimental data set, previously published Illumina sequencing data from the NA12878 individual that was aligned using BWA was considered. LUMPY considered paired-end (*pe*) and split read (*sr*) alignments for NA12878, LUMPY prior considered paired-end alignments and split read alignments for NA12878 and 1000 Genomes variants as prior evidence (*prior*), LUMPY trio considered paired-end alignments and split read alignments for NA12878 and her parents NA12891 and NA12892. DELLY considered paired-end (*pe*) and split-read (*sr*) alignments, GASVPro considered paired-end alignments and read-depth (*rd*), and PINDEL considered split-read (*sr*) alignments. Sensitivity and FDR was estimated using non-overlapping validated deletions from Mills et al. [XX] as well as predictions made by at least two tools. **(a)** SV detection sensitivity and FDR for the four tools on a 5X coverage subsample of the original data. LUMPY was more sensitive than both GASVPro and PINDEL and had an either equivalent or better FDR. DELLY was slightly more sensitive than LUMPY, but also had a much higher FDR. The inclusion of prior evidence or parental genomes improves sensitivity without increasing false positives. **(b)** SV detection sensitivity and FDR for the four tools on the original 50X coverage data. LUMPY, DELLY, and PINDEL had similar sensitivity in the Mills et al. case, and in the extended truth set LUMPY and DELLY had the highest sensitivity. LUMPY had lower

FDR than all other tools in both cases. The effect of prior evidence and parental genomes is reduced in higher coverage data.

**Figure 5. ROC curves comparing deletion prediction performance in the NA12878 individual.** ROC curves are shown comparing the correctly predicted deletions made by LUMPY, GASVPro, DELLY, and PINDEL to the deletions in the standard Mill et al truth set, as well as the extended truth set which also includes the deletion predictions identified by two or more tools. As in Figure 4, prediction performance was measured with both 5X average genome coverage (A) and 50X average genome coverage (B). Each tool was given the same set of input alignments in BAM format and the SV calls made by each tool were compared to the truth sets using identical criteria (see Methods). Each point on a given tool's ROC curve represents a minimum evidence support threshold ranging from 4 to 11, inclusive for 5X coverage and 4 to 15, inclusive for 50X coverage. The curves are colored following the same convention described in Figure 4.

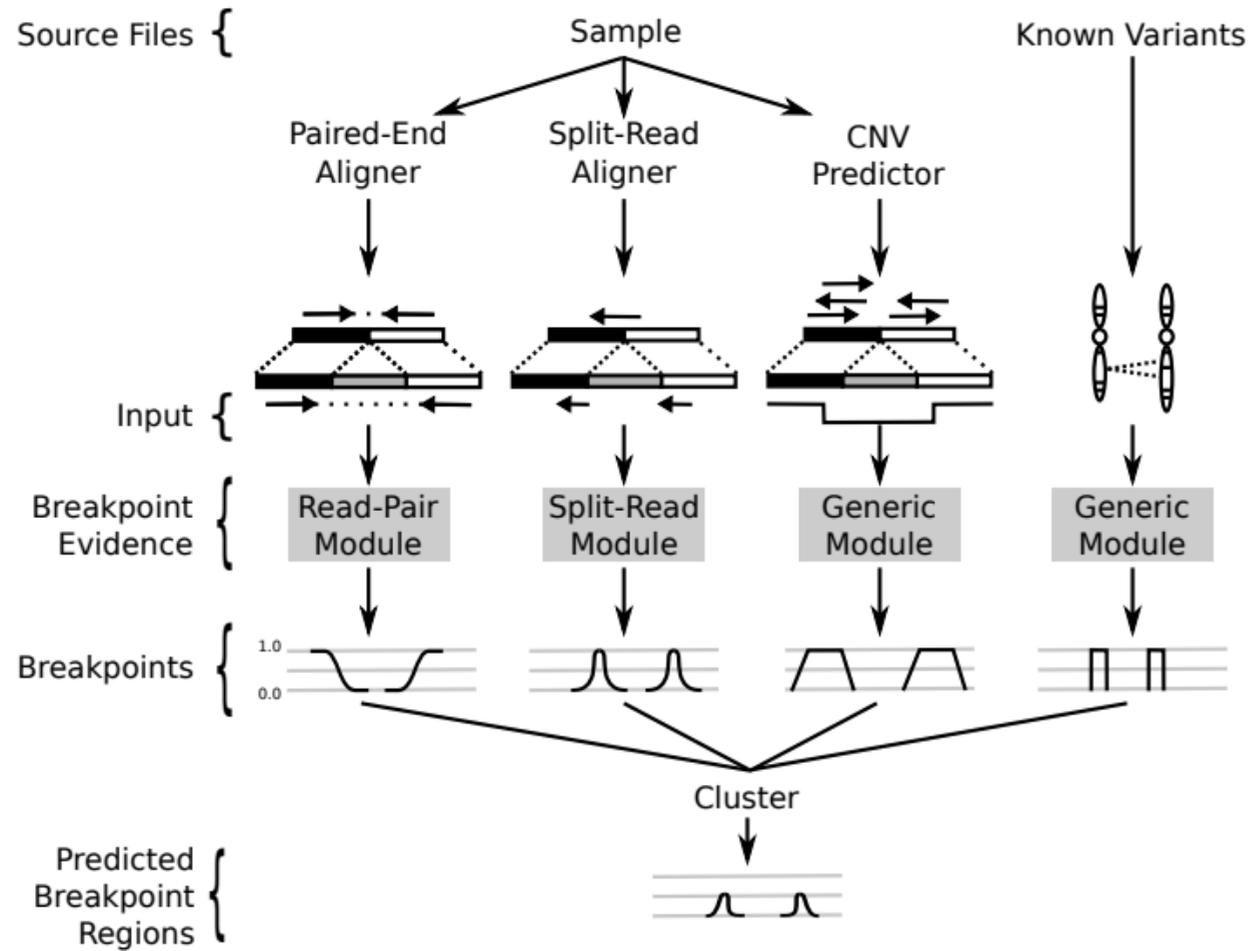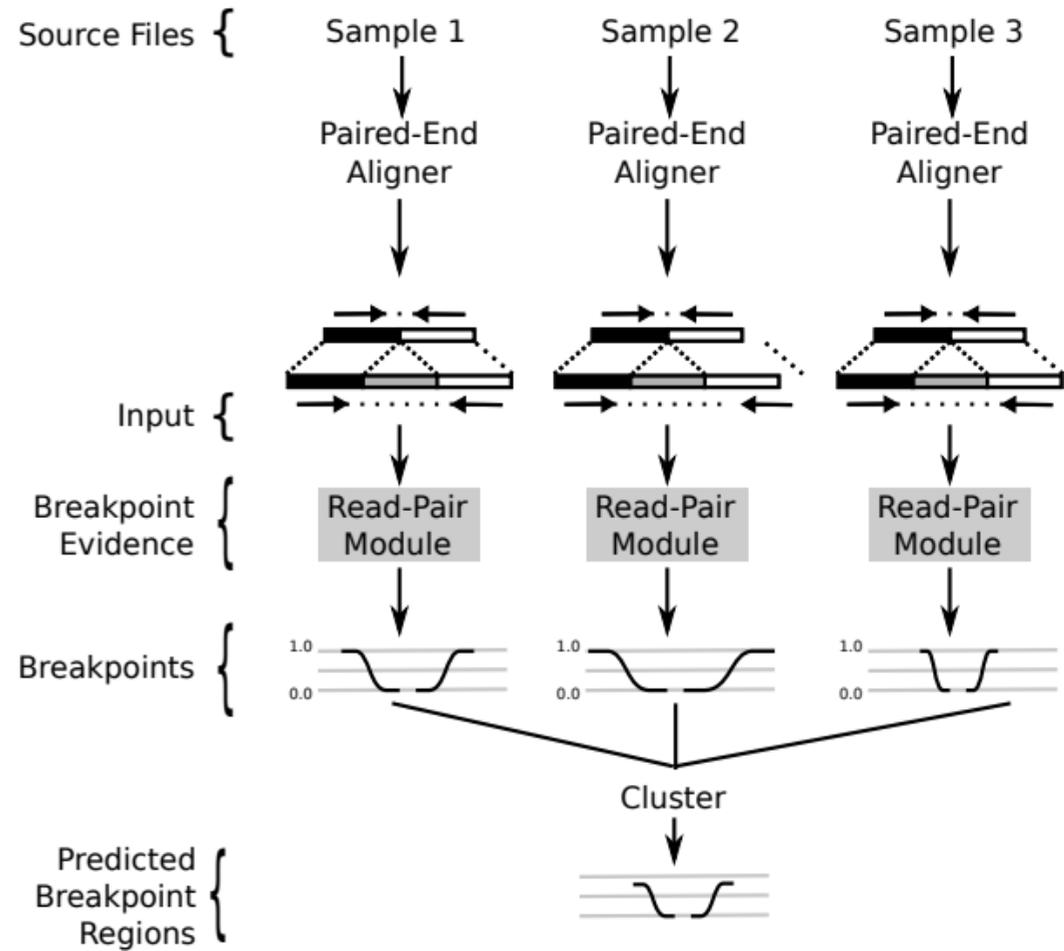

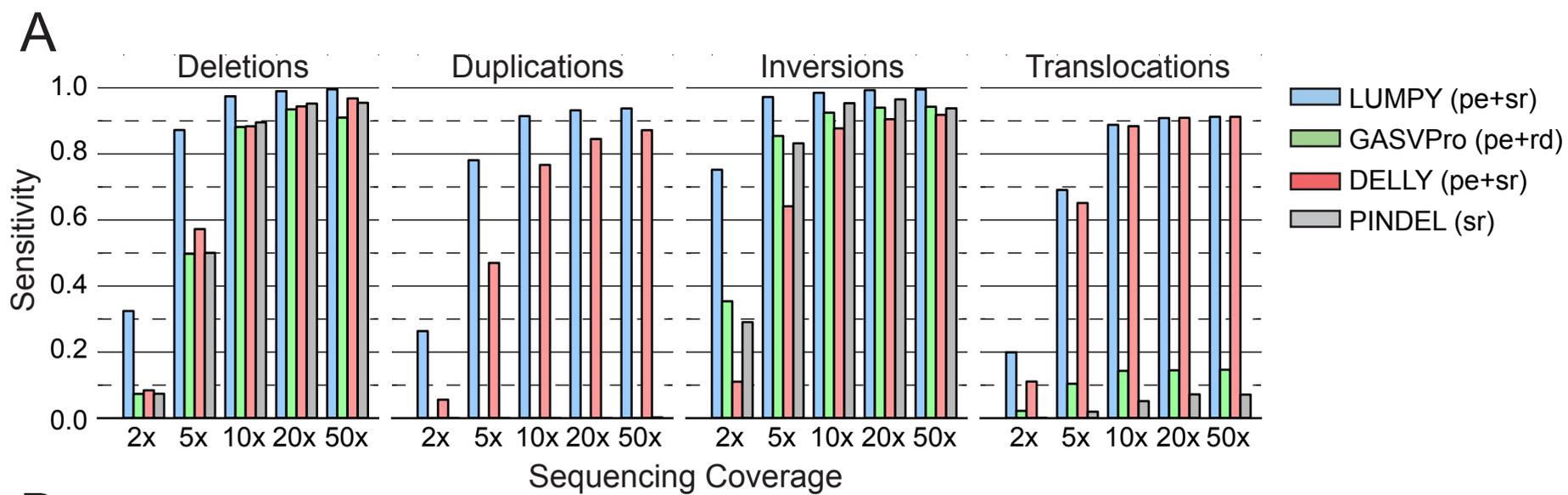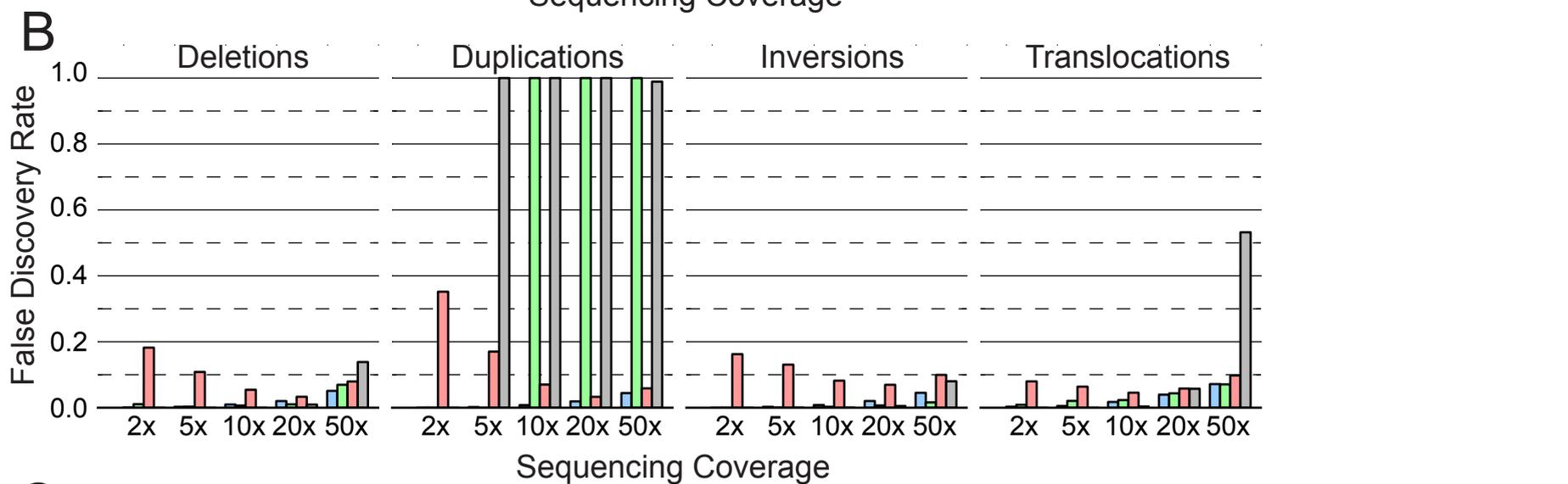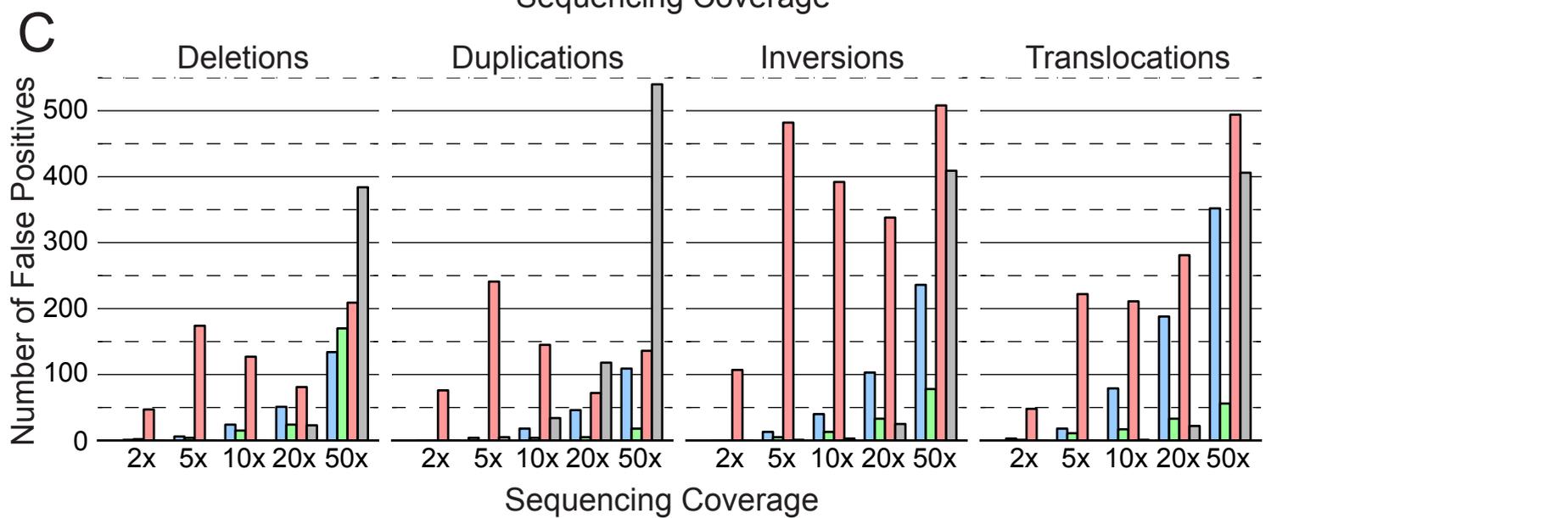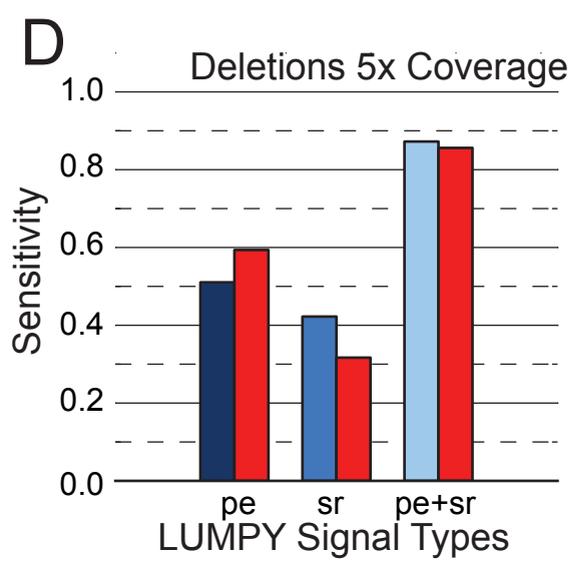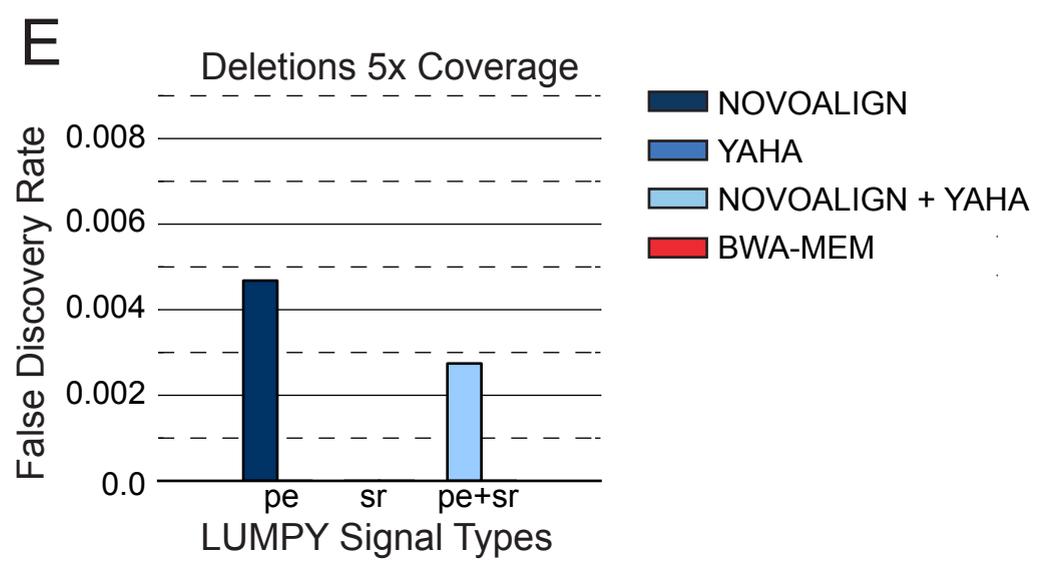

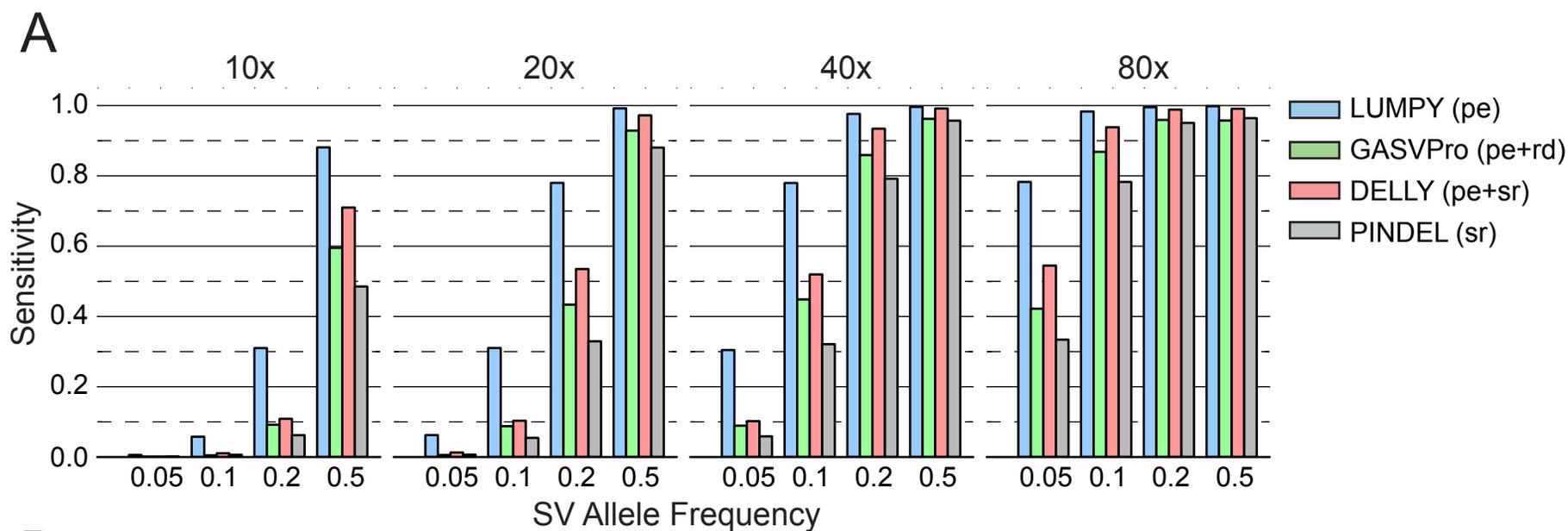
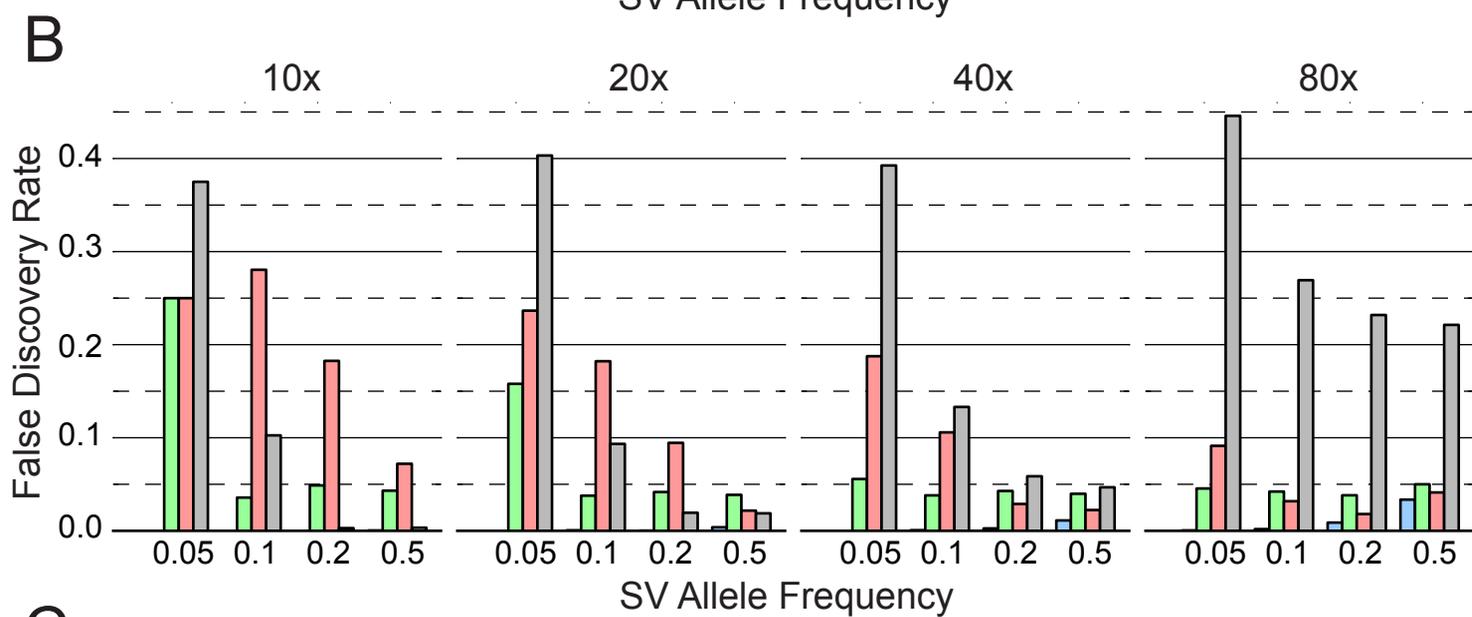
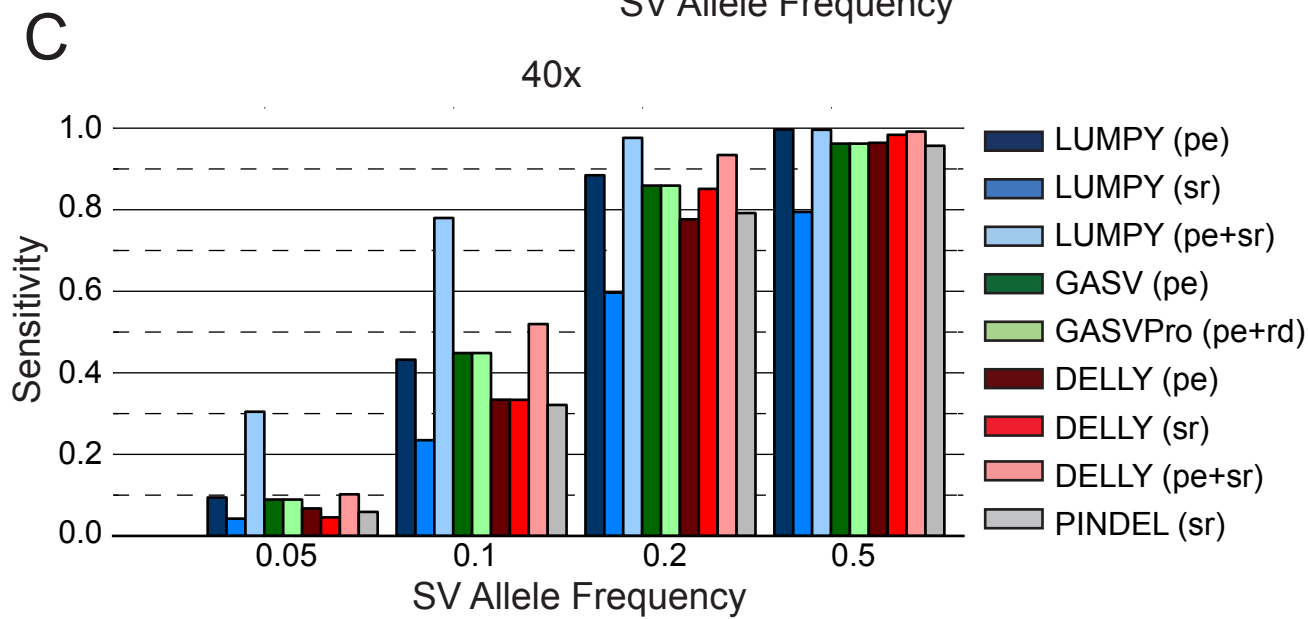

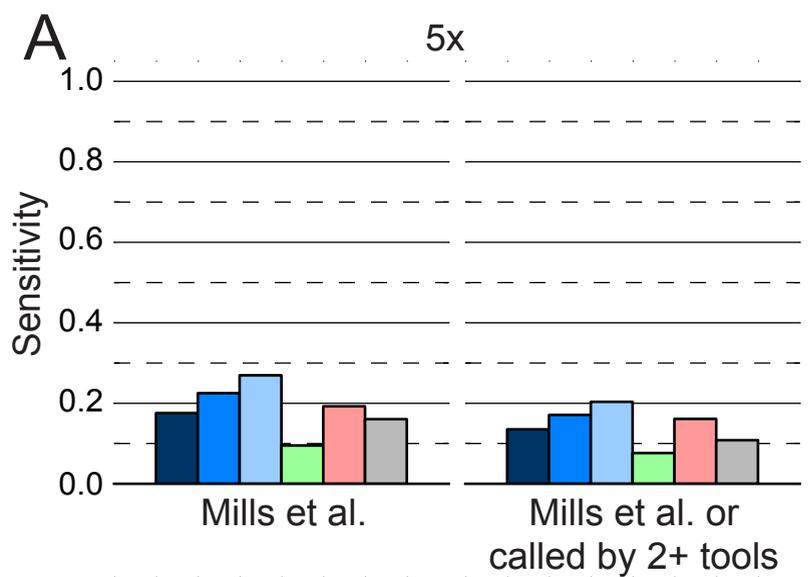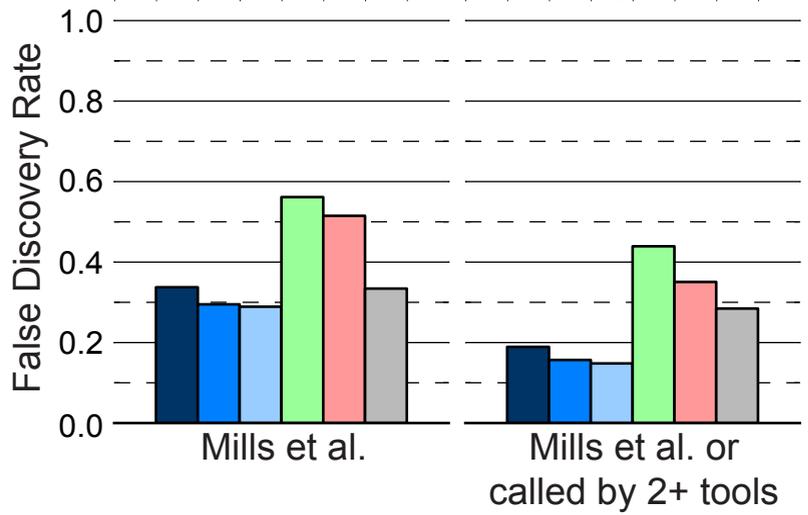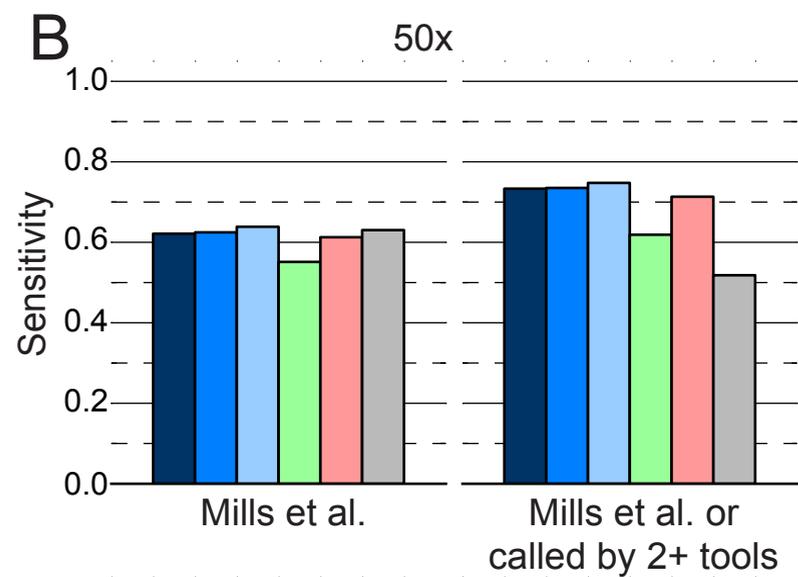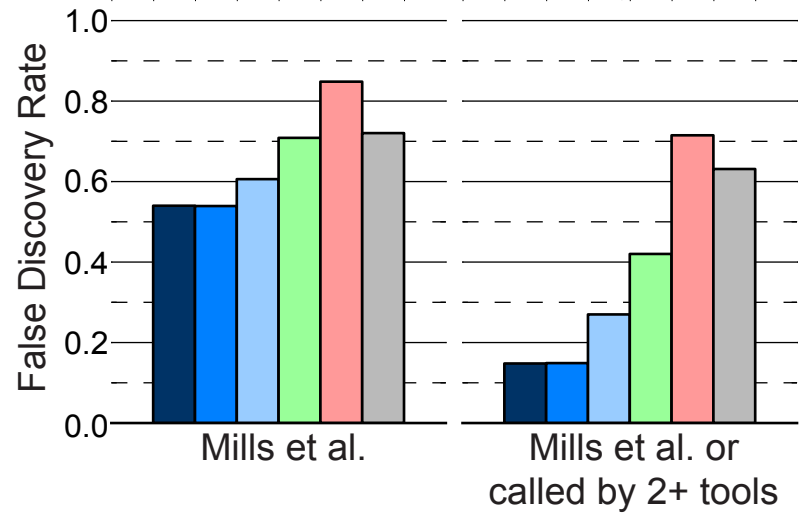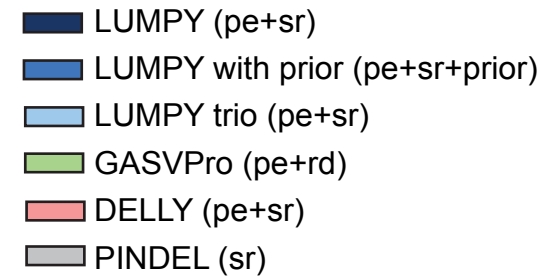

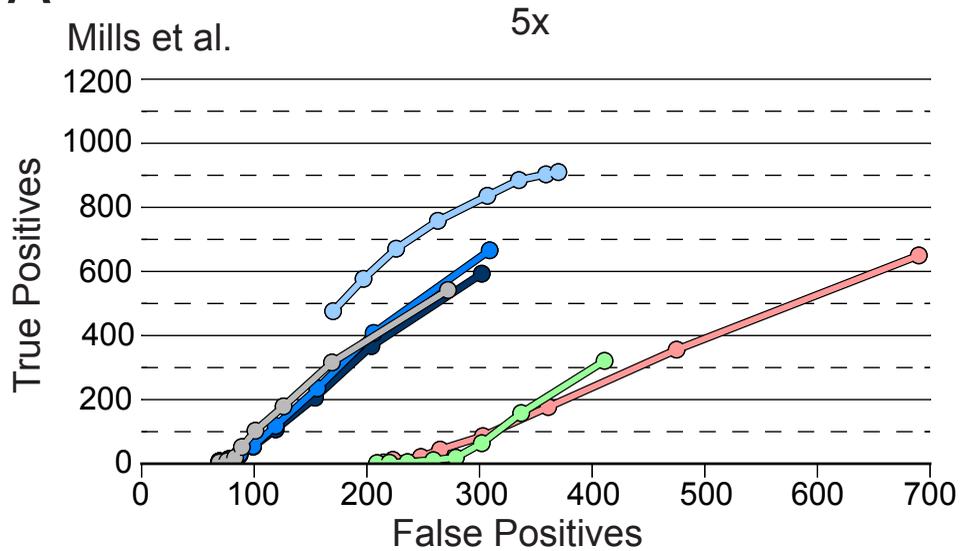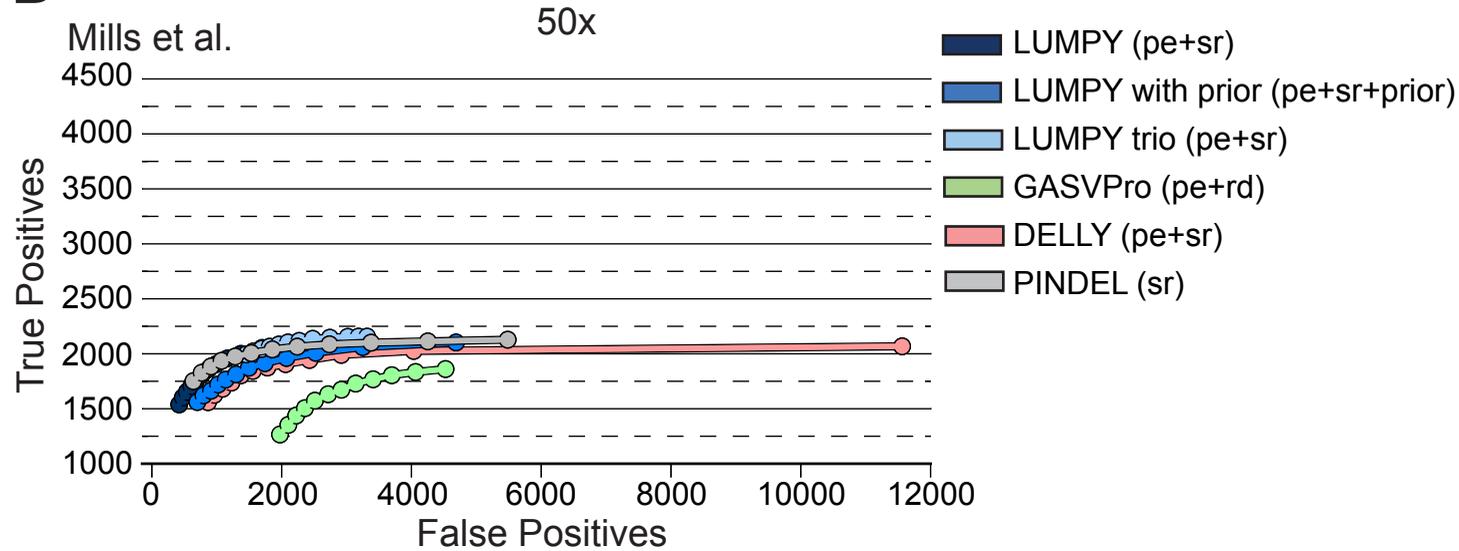